\documentclass[12pt,a4paper]{article}

\usepackage{epsfig}
\usepackage{amsmath,amsfonts,amssymb}
\usepackage{t1enc}
\usepackage{cite}

\newcommand{\pmiss}{\vec p\!\!\!\!\! \not \,\,\,\,}

\begin{document}
\vspace*{-3cm}
\begin{flushright}
hep-ph/0410068 \\
October 2004
\end{flushright}

\begin{center}
\begin{Large}
{\bf Sneutrino cascade decays $\boldsymbol{\tilde \nu_e \to e^- \tilde \chi_1^+
\to e^- f \bar f' \tilde \chi_1^0}$  \\ 
as a probe of chargino spin properties \\[0.3cm]
and CP violation}
\end{Large}

\vspace{0.5cm}
J. A. Aguilar--Saavedra \\[0.2cm] 
{\it Departamento de Física and CFTP, \\
  Instituto Superior Técnico, P-1049-001 Lisboa, Portugal} \\
\end{center}

\begin{abstract}
The decays $\tilde \nu_e \to e^- \tilde \chi_1^+$,
$\tilde \nu_e^* \to e^+ \tilde \chi_1^-$, when kinematically allowed,
constitute a source of 100\% polarised charginos.
We study the process $e^+ e^- \to \tilde \nu_e^* \tilde \nu_e \to e^+ \tilde
\chi_1^- \, e^- \tilde \chi_1^+$, with subsequent chargino decays
$\tilde \chi_1^- \to \bar \nu_\mu \mu^- \tilde \chi_1^0$,
$\tilde \chi_1^+ \to q \bar q' \tilde \chi_1^0$ or their charge conjugate
$\tilde \chi_1^- \to \bar q q' \tilde \chi_1^0$,
$\tilde \chi_1^+ \to \nu_\mu \mu^+ \tilde \chi_1^0$.
The kinematics of this process
allows the reconstruction of the sneutrino and chargino rest frames, and thus
a clean analysis of the angular distributions of the visible final state
products in these reference systems. Furthermore, a triple product CP asymmetry
can be built in $\tilde \chi_1^\pm$ hadronic decays, involving the chargino
spins and the momenta of the quark and antiquark.
With a good $c$ tagging efficiency, this CP asymmetry is quite sensitive to
the phase of the bino mass term $M_1$ and can test CP
violation in the neutralino sector.
\end{abstract}



\section{Introduction}
\label{sec:1}

An international linear $e^+ e^-$ collider (ILC) offers an ideal environment
for the study of
supersymmetry (SUSY) \cite{susy2,susy3}, if this theory is realised in nature. 
With adequate choices for the centre of mass (CM) energy and beam polarisations,
the various production processes simultaneously present in $e^+ e^-$
annihilation can be disentangled. It is then possible to measure sparticle
masses, couplings and mixings \cite{tdr}, allowing for the determination of the
Lagrangian parameters. The quantum numbers of the new particles must be
investigated as well, in order to confirm that they are the
superpartners of the Standard Model (SM) fields. In particular, it is crucial
to determine the spins of the SUSY particles. Spin-dependent effects serve
not only to deduce the particle spins but may also
be used to verify the predictions of the theory, using the input
from other experiments. It is also conceivable that spin distributions and
asymmetries, if precisely measured, can be used to improve the measurement of
some of the mixing parameters of the Lagrangian.

The existence of purely supersymmetric CP violation sources is other of the
aspects that require a thorough investigation. At present, all CP violation
observed in $K$ and $B$ oscillations and decay (with the possible exception of
the time-dependent CP asymmetry in $B_d^0 \to \phi K_S$, whose experimental
situation is still unclear) can be explained by
a single CP-violating phase in the Cabibbo-Kobayashi-Maskawa mixing matrix.
However, the CP
breaking induced by this phase cannot account for the observed baryon asymmetry
in the universe. The most general Lagrangian of the minimal supersymmetric
Standard Model (MSSM) contains a large number of CP violation sources,
which can potentially yield observable effects at low
and high energies. Within the neutralino sector, CP-violating
phases can appear in the bino and Higgsino mass terms, $M_1$ and $\mu$,
respectively.
At low energies the phases $\phi_1$, $\phi_\mu$ of these two parameters
generally lead to unacceptably large SUSY contributions to electric dipole
moments (EDMs), if they are different from $0,\pi$.  Present constraints from
EDMs place strong restrictions on the
values of $\phi_1$ and $\phi_\mu$, but do not necessarily require that $M_1$ and
$\mu$ are real, and cancellations among the different SUSY contributions can
occur \cite{pr1,pr2,kane}. This possibility, and the need for further CP
violation sources beyond the SM, makes the investigation of CP breaking in the
neutralino sector a compelling task to be carried out at a linear collider.

In this paper we focus on the determination of the spin and
spin-related properties of sneutrinos and charginos, including
CP-violating spin asymmetries in chargino decays. We study
sneutrino pair production in $e^+ e^-$ annihilation at a CM energy of 800 GeV,
as proposed for an ILC upgrade. We concentrate in the channels
\begin{align}
e^+ e^- & \to \tilde \nu_e^* \tilde \nu_e \to e^+ \tilde \chi_1^- \,
e^- \tilde \chi_1^+ \to e^+ \bar \nu_\mu \mu^- \tilde \chi_1^0 \,
e^- q \bar q' \tilde \chi_1^0 \nonumber \,,
  \nonumber \\
e^+ e^- & \to \tilde \nu_e^* \tilde \nu_e \to e^+ \tilde \chi_1^- \,
e^- \tilde \chi_1^+ \to e^+ \bar q q' \tilde \chi_1^0 \,
e^- \nu_\mu \mu^+ \tilde \chi_1^0 \,,
\label{ec:1}
\end{align}
with $q=u,c$, $q'=d,s$.
This process has three advantages for our purposes: ({\em i\/}) the charginos
produced are 100\% polarised, having positive helicity in the decay 
$\tilde \nu_e^* \to e^+ \tilde \chi_1^-$ and negative helicity in
$\tilde \nu_e \to e^- \tilde \chi_1^+$; ({\em ii\/}) the kinematics
allows the determination of the sneutrino and chargino rest frames, and
then the study of angular distributions in these reference systems;
({\em iii\/}) its cross section is large, and backgrounds with 5 energetic
final state particles plus large missing energy and momentum are small. Muon
sneutrinos are also a source of polarised charginos, but the cross section for
$\tilde \nu_\mu^* \tilde \nu_\mu$ production is much smaller.
Beam polarisations $P_{e^-} = -0.8$, $P_{e^+} = 0.6$ can be used to increase
the signal cross sections but they do not affect the angular distributions and
CP asymmetries studied. 

For definiteness, we consider a SUSY scenario like SPS1a in Ref.~\cite{sps}
but with a heavier slepton spectrum (and nonzero phases $\phi_1$, $\phi_\mu$),
so that sneutrino decays to charginos are kinematically allowed.
The analysis of $e^-$, $e^+$ angular distributions in the $\tilde \nu_e$,
$\tilde \nu_e^*$ rest frames provides a strong indication that sneutrinos
are scalar particles and charginos have spin $1/2$. The angular distributions
of the
$\tilde \chi_1^-$ ($\tilde \chi_1^+$) decay products $\mu^-$, $\bar q$, $q'$
($\mu^+$, $q$, $\bar q'$) with respect
to the $\tilde \chi_1^-$ ($\tilde \chi_1^+$) spin can be precisely measured as
well. Furthermore, summing $\tilde \chi_1^-$ and $\tilde \chi_1^+$
hadronic decays, a CP asymmetry based on the triple product
$\vec s_\pm \cdot (\vec p_{\bar q_1} \times \vec p_{q_2})$ can be built, where
$\vec s_\pm$ is
the spin direction of the $\tilde \chi_1^-$ or $\tilde \chi_1^+$, and
$\vec p_{\bar q_1}$, $\vec p_{q_2}$ are
the three-momenta of the antiquark and quark resulting from its decay,
respectively, in the
$\tilde \chi_1^\pm$ rest frame. A good $c$ tagging efficiency, as
expected for a future linear collider,
is essential for the determination of quark angular distributions and the
CP asymmetry. For the latter it is necessary to
distinguish between the quark and antiquark produced in the decay. This can be
done combining $c$ tagging and the knowledge of the final state muon charge.

We note that in chargino pair production $e^+ e^- \to
\tilde \chi_1^+ \tilde \chi_1^-$ the charginos are polarised as well
\cite{abdel}, allowing for the measurement of decay angular distributions and
CP-violating asymmetries.
However, in this process the momenta of the decaying charginos cannot
be determined with kinematical constraints, and thus the analysis of angular
distributions is less clean. Moreover, the backgrounds for
$\tilde \chi_1^+ \tilde \chi_1^-$ in the semileptonic decay channel are huge,
being $e^+ e^- \to W^+ W^- \to \ell^\pm \nu jj$, $\ell=e,\mu$ the most
troublesome,
with a cross section (including ISR and beamstrahlung corrections) of
approximately 3.5 pb at CM energies of 500 and 800 GeV, for $P_{e^+}=0.6$,
$P_{e^-} = -0.8$. At any rate, in SUSY scenarios where sneutrino decays
to charginos are not kinematically allowed,
$\tilde \chi_1^+ \tilde \chi_1^-$ production seems the best place for the
study of chargino spin properties and CP asymmetries in chargino decays.

This paper is organised as follows. In section~\ref{sec:2} we discuss sneutrino
production and decay, focusing on the features most relevant for our
work, and fix the SUSY scenario used. The procedure for the Monte
Carlo calculation of the processes and reconstruction of the signals is
outlined in section~\ref{sec:3}. In section~\ref{sec:4} we introduce the
different distributions analysed and present our numerical
results. In section~\ref{sec:5} we compare
with results obtained from other processes and draw our conclusions.

\section{Production and decay of sneutrino pairs}
\label{sec:2}

\begin{figure}[htb]
\begin{center}
\begin{tabular}{ccc}
\mbox{\epsfig{file=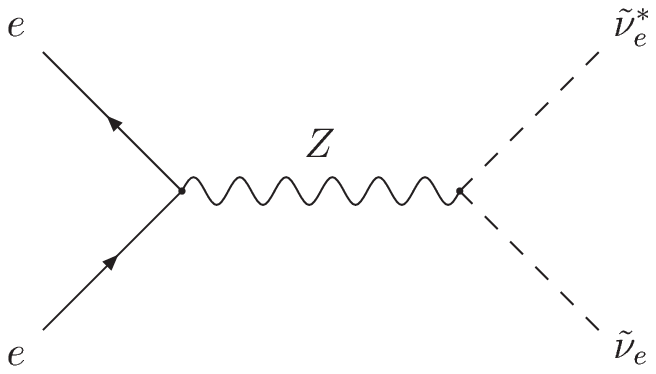,width=3.8cm,clip=}} & \hspace*{5mm} & 
\mbox{\epsfig{file=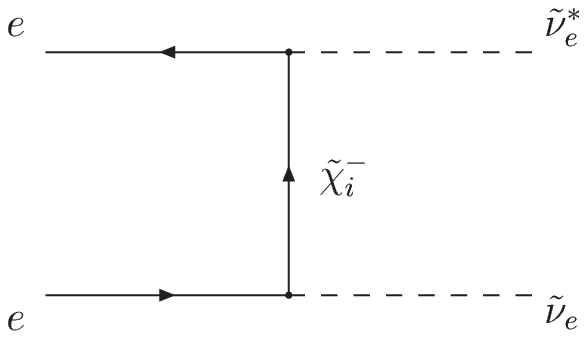,width=3.8cm,clip=}} 
\end{tabular}
\caption{Feynman diagrams for $\tilde \nu_e^* \tilde \nu_e$ production
in $e^+ e^-$ annihilation.}
\label{fig:1}
\end{center}
\end{figure}

\begin{figure}[htb]
\begin{center}
\begin{tabular}{ccc}
\epsfig{file=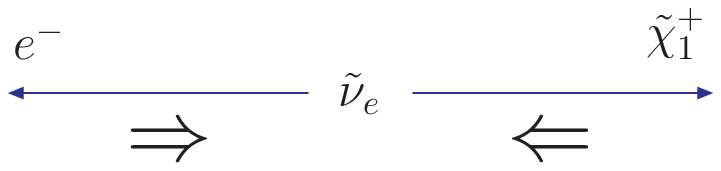,width=5cm,clip=} & ~~~~ & 
\epsfig{file=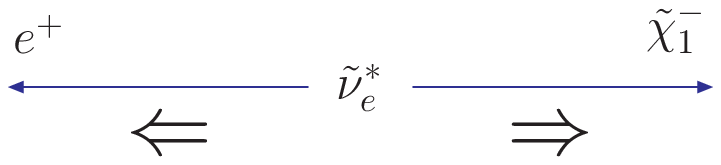,width=5cm,clip=} \\
(a) & & (b)
\end{tabular}
\caption{Helicity of the charginos produced in $\tilde \nu_e$ and $\tilde
\nu_e^*$ decays.}
\label{fig:2}
\end{center}
\end{figure}

Sneutrino pairs are produced in $e^+ e^-$ collisions through the Feynman
diagrams depicted in Fig.~\ref{fig:1}. Their two-body decays
$\tilde \nu_e \to e^- \tilde \chi_1^+$, $\tilde \nu_e^* \to e^+ \tilde \chi_1^-$
are mediated by the $\tilde \nu_e e \tilde \chi_1^-$ vertex which, neglecting
the electron Yukawa coupling, is given by
\begin{eqnarray}
\mathcal{L}_{\tilde \nu_e e \tilde \chi_1^-} & = & -g V_{11} \,
 \bar e \, P_R \, \tilde \chi_1^- \; \tilde \nu_e  -g V_{11}^* \,
\overline{\tilde \chi_1^-} \, P_L \, e \; \tilde \nu_e^* \,,
\label{ec:2}
\end{eqnarray}
with $V^\dagger$ the $2 \times 2$ unitary matrix diagonalising the chargino mass
matrix by the right (the interactions and notation used can be found in
Refs.~\cite{npb,x1x2}, and follow the conventions of Ref.~\cite{romao}). By
inspection of the Lagrangian it is easily seen
that in the decay $\tilde \nu_e \to e^- \tilde \chi_1^+$, mediated by 
the first term in Eq.~(\ref{ec:2}), the produced electron has negative chirality
(and thus negative helicity, neglecting the electron mass) and the chargino
has positive chirality. Since sneutrinos are spinless particles, angular
momentum conservation in the $\tilde \nu_e$ rest frame implies that the
$\tilde \chi_1^+$ must have negative
helicity as well, as depicted schematically in Fig.~\ref{fig:2} (a).
For $\tilde \nu_e^* \to e^+ \tilde \chi_1^-$, the positron has negative
chirality, and thus positive helicity; therefore, the $\tilde \chi_1^-$ has
positive helicity in the $\tilde \nu_e^*$ rest frame, as shown in 
Fig.~\ref{fig:2} (b).

\noindent

\begin{figure}[htb]
\begin{center}
\begin{tabular}{ccccc}
\mbox{\epsfig{file=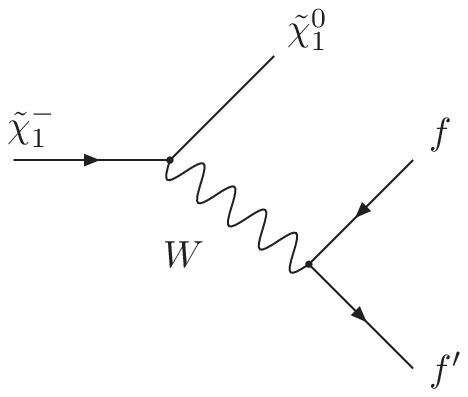,width=3cm,clip=}} & \hspace*{5mm} & 
\mbox{\epsfig{file=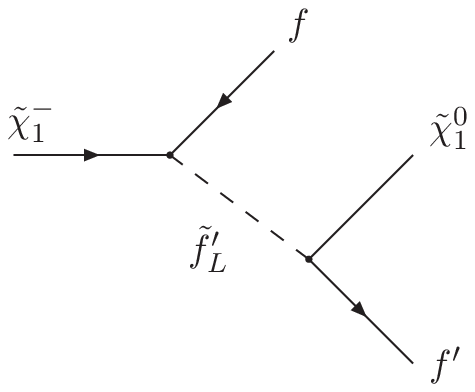,width=3cm,clip=}} & \hspace*{5mm} & 
\mbox{\epsfig{file=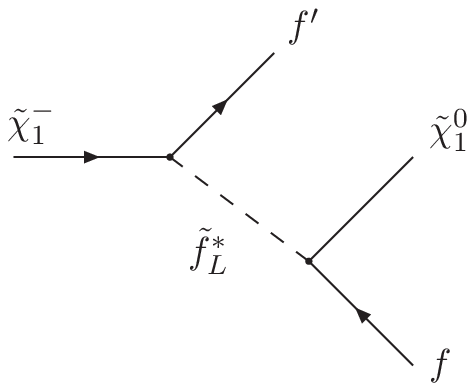,width=3cm,clip=}} \\
(a) & & (b) & & (c)
\end{tabular}
\caption{Feynman diagrams for the decay $\tilde \chi_1^- \to 
\bar f f' \tilde \chi_1^0$, for light fermions
$\bar f f' = \bar \nu_e e^-, \bar \nu_\mu \mu^-, \bar u d,\bar c s$ and
negligible
$\tilde f'_L-\tilde f'_R$, $\tilde f_L-\tilde f_R$ mixing. For $f=\nu_\ell$,
in diagram (c) $\tilde f_L^* = \tilde \nu_\ell^*$ is exchanged.}
\label{fig:3}
\end{center}
\end{figure}

The decay of $\tilde \chi_1^-$ to light fermions $\bar f,f'$ and a neutralino
$\tilde \chi_1^0$ is mediated by the diagrams in Fig.~\ref{fig:3}, with
similar diagrams for $\tilde \chi_1^+$ decay. We study final states in which
one of the charginos decays leptonically and the other hadronically. In leptonic
decays we restrict ourselves to $\bar f f' = \bar \nu_\mu \mu^-$
($f \bar f' = \nu_\mu \mu^+$). The reason is that in decays to $e^\pm$ and a
neutrino, the
presence of an additional $e^+ e^-$ pair from sneutrino decays difficults
the reconstruction of the final state, while in decays to $\tau^\pm$ the
momentum of the charged lepton cannot be directly measured. We sum all hadronic
decays $\bar f f' = \bar u d,\bar c s$ ($f \bar f' = u \bar d,c \bar s$).

We choose a SUSY scenario similar to SPS1a but with a heavier sfermion spectrum
and complex phases $\phi_1$, $\phi_\mu$. The low-energy parameters
most important for our analysis are collected in Table~\ref{tab:1}. 
For vanishing $\phi_1$, $\phi_\mu$, they approximately correspond to
$m_{1/2} = 250$ GeV,
$m_{\tilde E} = m_{\tilde L} = m_{H_i} = 200$ GeV, $A_E = -200$ GeV at the
unification
scale, and $\tan \beta = 10$. We use {\tt SPheno} \cite{spheno} to calculate
sparticle masses, mixings and some decay widths.
Neutralino and chargino masses slightly depend on $\phi_1$, $\phi_\mu$;
for $\phi_1 = \phi_\mu = 0$ they are $m_{\tilde \chi_1^0} = 99$ GeV,
$m_{\tilde \chi_1^-} = 178$ GeV, $m_{\tilde \chi_2^-} = 401$ GeV.
The relevant branching ratios
(taking $\phi_1 = \phi_\mu = 0$) are
$\mathrm{Br}(\tilde \nu_e \to e^- \, \tilde \chi_1^+) = 0.52$,
$\mathrm{Br}(\tilde \chi_1^- \to \bar \nu_\mu \mu^- \tilde \chi_1^0) = 0.10$,
$\mathrm{Br}(\tilde \chi_1^- \to \bar q q' \tilde \chi_1^0) = 0.34$, with the
same rates for the charge-conjugate processes.

\begin{table}[htb]
\begin{center}
\begin{tabular}{ccc}
Parameter & ~ & Value \\
\hline
$M_1$ & & 102.0 $e^{i \phi_1}$ \\
$M_2$ & & 192.0 \\
$\mu$ & & 377.5 $e^{i \phi_\mu}$ \\
$\tan \beta$ & & 10 \\
$m_{\tilde \nu_e}$ & & 252.4 \\
$m_{\tilde \mu_L}$ & & 264.5 \\
$m_{\tilde u_L},m_{\tilde c_L}$ & & 571.5 \\
$m_{\tilde d_L},m_{\tilde s_L}$ & & 577.0
\end{tabular}
\caption{Low-energy parameters (at the scale $M_Z$) for the SUSY scenario used.
The dimensionful parameters are in GeV.
\label{tab:1}}
\end{center}
\end{table}

In the scenario selected the three-body decays $\tilde \chi_1^- \to f \bar f'
\tilde \chi_1^0$ are mediated by off-shell intermediate particles. Still, they
are dominated by $W$ exchange, diagram (a) in Fig.~\ref{fig:3}, due to the large
sfermion masses. In
$\tilde \chi_1^- \to \bar \nu_\mu \mu^- \tilde \chi_1^0$ the second diagram
in importance is sneutrino exchange in diagram \ref{fig:3} (c), but its
contribution to the partial width is 13 times smaller than the one from $W$
exchange. For hadronic decays,
diagrams involving squarks are even more suppressed, and their contribution
has a relative magnitude of $2 \times 10^{-4}$ with respect to the $W$
diagram. Therefore, many
characteristics of this scenario are shared with scenarios with a heavier
chargino spectrum,
in which the two-body decay $\tilde \chi_1^- \to W^- \tilde \chi_1^0$ 
is possible and diagrams (b) and (c) are negligible both for leptonic and
hadronic final states.

For values of $\phi_1$, $\phi_\mu$ different from $0,\pi$, CP is violated
in the neutralino and chargino sectors, leading to large SUSY
contributions to EDMs. If the squark spectrum (which does
not play any role in our analysis, as seen from the previous discussion) is
heavy enough, experimental limits on the neutron and Mercury EDMs can be
satisfied. On the other hand, for the selectron and sneutrino masses
under consideration, the experimental bound on the electron EDM $d_e$ severely
constrains the allowed region in the $(\phi_1,\phi_\mu)$ plane. Using
the expressions for $d_e$ in Ref.~\cite{arnowitt} it is found that, for each
$\phi_1$ between 0 and $2 \pi$, in this scenario there exist two narrow
intervals for $\phi_\mu$, one with values $\phi_\mu \sim 0$ and the other with
values $\phi_\mu \sim \pi$, in which the neutralino and chargino contributions
to $d_e$ partially cancel, resulting in a value
compatible with the experimental limit $d_e^\mathrm{\,exp} = (0.079 \pm 0.074)
\times 10^{-26} ~ e$ cm \cite{pdb}. For instance, for $\phi_1 = \pi/2$ the
phase $\phi_\mu$ can be $\phi_\mu \simeq -0.12$ or $\phi_\mu \simeq 3.21$.
Several representative examples of these pairs of phases allowed by EDM
constraints are collected in Table~\ref{tab:pairs}. 
We can see that in principle it is possible to have any phase $\phi_1$,
though with a strong correlation with $\phi_\mu$.
If $\phi_1$ and $\phi_\mu$ are experimentally found to be
non-vanishing, a satisfactory explanation will be necessary for this
correlation, which apparently would be a
``fine tuning'' of their values \cite{pr3}.

\begin{table}[htb]
\begin{center}
\begin{tabular}{ccccc}
$\phi_1$ & $\phi_\mu$ & & $\phi_1$ & $\phi_\mu$ \\
\hline
0 & 0 & & $\pi$ & 0 \\
$\pi/8$ & -0.0476 & ~ & $7\pi/8$ & -0.0454 \\
$\pi/4$ & -0.0876 &   & $3\pi/4$ & -0.0845 \\
$3\pi/8$ & -0.1136 &  & $5\pi/8$ & -0.1114 \\
$\pi/2$ & -0.1218
\end{tabular}
\caption{Examples of approximate phases $\phi_1$, $\phi_\mu$ for which
the pairs $(\phi_1,\phi_\mu)$, $(-\phi_1,-\phi_\mu)$
are allowed by EDM constraints.}
\label{tab:pairs}
\end{center}
\end{table}

\section{Generation and reconstruction of the signals}
\label{sec:3}

The matrix element of the resonant processes in Eq.~(\ref{ec:1}) are
calculated using {\tt HELAS} \cite{helas}, including all spin correlations
and finite width effects. The
relevant terms of the Lagrangian can be found in Refs.~\cite{npb,x1x2}. 
We assume a CM energy of 800 GeV, with electron polarisation $P_{e^-} = -0.8$
and positron polarisation $P_{e^+} = 0.6$. Beam polarisation has no effect on
the angular distributions and asymmetries studied but increases the signal cross
section. The luminosity is taken as 534 fb$^{-1}$ per year \cite{lum}.
In our calculations we take into account the effects of initial state radiation
(ISR) \cite{isr} and beamstrahlung \cite{peskin,BS2}. For the design luminosity
at 800 GeV we use the parameters $\Upsilon = 0.09$, $N = 1.51$ \cite{lum}.
The actual expressions for ISR and beamstrahlung used in our calculation are
collected  in Ref.~\cite{npb}. We also include a beam energy spread of 1\%.

We simulate the calorimeter and tracking resolution of the detector by
performing a Gaussian smearing of the energies of electrons ($e$), muons ($\mu$)
and jets ($j$), using the specifications in Ref.~\cite{tesla2},
\begin{equation}
\frac{\Delta E^e}{E^e} = \frac{10\%}{\sqrt{E^e}} \oplus 1 \% \;, \quad
\frac{\Delta E^\mu}{E^\mu} = 0.02 \% \, E^\mu \;, \quad
\frac{\Delta E^j}{E^j} = \frac{50\%}{\sqrt{E^j}} \oplus 4 \% \;,
\end{equation}
where the two terms are added in quadrature and the energies are in GeV.
We apply kinematical cuts on transverse momenta, $p_T \geq 10$ GeV, and
pseudorapidities $|\eta| \leq 2.5$, the latter corresponding to polar angles
$10^\circ \leq \theta \leq 170^\circ$. We also reject events in which the
leptons or jets are not isolated, requiring a ``lego-plot'' separation
$\Delta R = \sqrt{\Delta \eta^2+\Delta \phi^2} \geq 0.4$.
For the Monte Carlo integration in 10-body phase space we use
{\tt RAMBO} \cite{rambo}.

For the precise measurement of angular distributions it is crucial to
reconstruct accurately the final state. This is more difficult when
several particles escape detection, as it happens in our case. Since we are not
interested in the mass distributions we use as input all the sparticle masses
involved, which we assume measured in other processes \cite{tdr}. Let us label
the
electron and positron 4-momenta as $p_{e^-}$, $p_{e^+}$, respectively, and
the  momenta of the ``visible'' chargino decay products (the $\mu^\pm$ for
the leptonic decay and the quark-antiquark pair in the hadronic decay) as
$p_{V^+}$, $p_{V^-}$. The unknown momenta of the ``invisible'' chargino decay
products (the $\nu \tilde \chi_1^0$ pair in leptonic decays and the neutralino
in hadronic decays) are $p_{I^+}$, $p_{I^-}$ (8 unknowns). From four-momentum
conservation and the kinematics of the process we have the relations
\begin{align}
E_{I^+} + E_{I^-} & = E_\mathrm{CM} - E_{e^-} - E_{e^+}
- E_{V^+} - E_{V^-} \,, \nonumber \\
\vec p_{I^-} + \vec p_{I^+} & = \pmiss \,, \nonumber \\
(p_{V^+} + p_{I^+})^2 & = m_{\tilde \chi_1^-}^2 \,, \nonumber \\
(p_{V^-} + p_{I^-})^2 & = m_{\tilde \chi_1^-}^2 \,, \nonumber \\
(p_{e^-} + p_{V^+} + p_{I^+})^2 & = m_{\tilde \nu_e}^2 \,, \nonumber \\
(p_{e^+} + p_{V^-} + p_{I^-})^2 & = m_{\tilde \nu_e}^2 \,,
\label{ec:rec}
\end{align}
where $E_\mathrm{CM}$ is the CM energy and $\pmiss$ the missing momentum.
These 8 equations are at most quadratic in the unknown momenta $p_{I^+}$,
$p_{I^-}$. With a little algebra, they can be written in the form of 7 linear
plus one quadratic equation, which can be solved yielding 2 possible solutions
for the unknown momenta (note that in leptonic decays only the sum of the
neutrino and neutralino momenta can be determined). In order
to select the correct one we use the additional constraint that either
$p_{I^+}^2 = m_{\tilde \chi_1^0}^2$ or $p_{I^-}^2 = m_{\tilde \chi_1^0}^2$
depending on which chargino decays hadronically (this is decided event by
event depending on the charge of the muon). With $p_{I^+}$, $p_{I^-}$ obtained
from the reconstruction
process, the momenta of the two charginos and the two sneutrinos can be
determined, as well as their respective rest frames.

We note that ISR, beamstrahlung, particle width effects and detector
resolution degrade the determination of $p_{I^+}$, $p_{I^-}$, being ISR and
beamstrahlung the most troublesome. Detector resolution affects the measurement
of charged lepton and jet momenta, while ISR and beamstrahlung modify the beam
energies and thus reduce the CM energy, causing also that the CM and laboratory
frames do not coincide. Moreover, the last four of Eqs.~(\ref{ec:rec}) only hold
for strictly on-shell sneutrinos and charginos.
Due to these effects, Eqs.~(\ref{ec:rec}) sometimes do not have a real
solution, {\em i.e.} the discriminant of the quadratic equation mentioned
above is negative. In such case, we force the system to have a real solution by
setting the discriminant to zero, what has the consequence that the
solutions do not fulfill Eqs.~(\ref{ec:rec}) for the input values
$m_{\tilde \chi_1^-}$, $m_{\tilde \nu_e}$ used but rather for other
(sometimes very different) ones.

The determination of the unknown momenta is done as follows.
In order to partially take
into account the decrease in the CM energy we replace $E_\mathrm{CM}$ in
Eq.~(\ref{ec:rec}) by an ``effective'' CM energy $E_\mathrm{CM}^\mathrm{eff}$.
We try the reconstruction of the unknown momenta for different energies
$E_\mathrm{CM}^\mathrm{eff} \leq E_\mathrm{CM}$ in decreasing order, choosing
the one which gives
reconstructed sneutrino, chargino and neutralino masses closest to their true
values. In case that different effective energies yield equal results for the
reconstructed masses, we choose the largest one. If
the event does not reasonably fit into the kinematics assumed for the process,
it is discarded.

For the analysis of some observables we take advantage of the good $c$
tagging capability that it is expected at a future linear collider. We use a
$c$ tagging efficiency of 50\%
and a mistag rate of 0.2\% \cite{ctag}. The identification of $c$ against
$\bar c$, when needed, can be indirectly done looking at the charge of the
muon produced. In final states with $\mu^-$, the chargino decaying hadronically
is $\tilde \chi_1^+$, thus the tagged jet corresponds to a $c$ quark and the
other one to a $s$ antiquark. By the
same argument, for $\mu^+$ final states the tagged jet corresponds to a $c$
antiquark and the other one to a $s$ quark.

The backgrounds to the processes in Eq.~(\ref{ec:1}) are rather small.
Other SUSY particle production processes leading to a final state of
$e^+ e^- \mu^\pm jj$ plus large
missing energy and momentum are for instance
$e^+ e^- \to \tilde \chi_1^\pm \tilde \chi_2^\mp \to \tilde \chi_1^\pm \; Z
\tilde \chi_1^\mp$,
$e^+ e^- \to \tilde \chi_1^\pm \tilde \chi_2^\mp \to \tilde \chi_1^\pm \; W^\mp 
\tilde \chi_2^0$,
$e^+ e^- \to  \tilde \chi_2^0 \tilde \chi_{3,4}^0 \to \tilde \chi_2^0
\, \tilde \chi_1^\pm \tilde W^\mp$, with subsequent decays $Z \to e^+ e^-$,
$\tilde \chi_2^0 \to e^+ e^- \tilde \chi_1^0$ and hadronic or leptonic
$\tilde \chi_1^\pm$, $W^\pm$
decays giving a muon, two jets, a neutrino and a neutralino. The total
cross section of the three processes amounts to 0.1 fb. More important is the
SM background $e^+ e^- \to e^+ e^- \mu^- \bar \nu q \bar q'$ (which includes
on-shell $Z W^+ W^-$ production) plus its charge
conjugate, with a cross section around 4 fb \cite{lusifer}. All these
backgrounds are expected to be considerably reduced by the
signal reconstruction process, which requires that the kinematics of the
events is compatible with sneutrino pair production. Six fermion production
may be further suppressed with a kinematical cut requiring that the invariant
mass of the $e^+ e^-$ pair is not consistent with $M_Z$.

\section{Results}
\label{sec:4}

We present our numerical results taking $\phi_1 = \phi_\mu = 0$ everywhere
except for the study of CP asymmetries. We collect in Table~\ref{tab:cs} the
total cross section for the
processes in Eq.~(\ref{ec:1}) (including ISR, beamstrahlung and beam spread
corrections), the cross section after ``detector'' cuts, its value including
also  reconstruction cuts, and finally requiring one $c$ tag as well. In the
latter case, only chargino decays to $\bar q q' = \bar c s$ ($q \bar q' = c \bar
s$) contribute in practice.

\begin{table}[htb]
\begin{center}
\begin{tabular}{lc}
& Cross section \\
\hline
Total & 17.56 \\
Detector & 11.05 \\
Reconstruction & 9.99 \\
$c$ tagging & 2.50
\end{tabular}
\caption{Total cross section (in fb) of the processes in Eq.~(\ref{ec:1})
before and after detector and reconstruction cuts, and including also $c$
tagging.}
\label{tab:cs}
\end{center}
\end{table}

We give our theoretical predictions for angular distributions
(calculated with sufficiently high Monte Carlo statistics) together with
a possible experimental result for one year of running (with an integrated
luminosity of 534 fb$^{-1}$). The latter is generated as follows: for each bin
$i$ we calculate the expected number of events in one year $\mu_i$, then the
``observed'' number of events $n_i$ in each bin is randomly obtained according
to a Poisson distribution with mean $\mu_i$. 
This procedure is done independently for each kinematical
distribution studied. In order to be not too optimistic in our results, we
present samples of possible experimental measurements in which the total number
of events is approximately $1 \, \sigma$ away from the theoretical expectation.

\subsection{Electron angular distributions}

\begin{figure}[htb]
\begin{center}
\epsfig{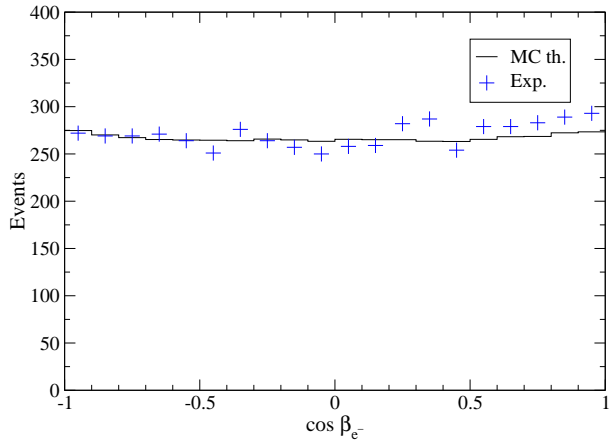}
\caption{Angular distribution of the $e^-$ with respect to an axis orthogonal
to the beam direction, in the $\tilde \nu_e$ rest frame.}
\label{fig:em}
\end{center}
\end{figure}

Sneutrino decays are predicted to be isotropic in their rest frame, as
corresponds to spinless particles. This fact can be experimentally tested in the
processes discussed here, with
the examination of the $e^-$ ($e^+$) angular distribution in the $\tilde \nu_e$
($\tilde \nu_e^*$) rest frame. We concentrate on the first case, defining
$\beta_{e^-}$, as the angle between the $e^-$ momentum (in $\tilde \nu_e$ rest
frame) and an axis orthogonal to the beam direction arbitrarily chosen.
In Fig.~\ref{fig:em} we show the dependence of the cross section on this angle.
The full line corresponds to the theoretical prediction, which
slightly deviates from a flat line due to detector and reconstruction cuts. The
points represent a possible experimental result.
Despite the small statistical fluctuations, it is clear that the result
corresponds to a flat distribution. Performing the analysis for three
orthogonal axes shows that the $\tilde
\nu_e$ decay is isotropic, what provides a strong indication that sneutrinos are
scalars and thus that charginos have half-integer spin (as implied by total
angular momentum conservation).

\subsection{Angular distributions of $\boldsymbol{\tilde \chi_1^-}$ decay
products}

\begin{figure}[htb]
\begin{center}
\epsfig{file=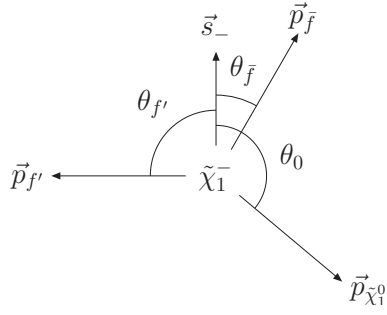,width=5cm,clip=}
\caption{Definition of the polar angles $\theta_{\bar f}$, $\theta_{f'}$ and
$\theta_0$ in the decay of $\tilde \chi_1^-$.}
\label{fig:5}
\end{center}
\end{figure}

The reconstruction of the chargino and sneutrino momenta allows for the
determination of
the chargino spin directions, which are $\vec s_- = \vec P_{\tilde \chi_1^-}$,
$\vec s_+ = - \vec P_{\tilde \chi_1^+}$ for $\tilde \chi_1^-$, $\tilde
\chi_1^+$, respectively, with $\vec P_{\tilde \chi_1^-}$,
$\vec P_{\tilde \chi_1^+}$ the chargino three-momenta in the $\tilde \nu_e$,
$\tilde \nu_e^*$ rest frames. The
knowledge of the chargino spin directions then makes it possible a precise
study of its polarised differential decay widths. The analytical expressions
for these quantities are rather lengthy \cite{djouadi}; however, the angular
distribution of a single decay product in the $\tilde \chi_1^-$ rest frame can
be cast in a compact form. Let us define $\theta_{\bar f}$, $\theta_{f'}$,
$\theta_0$ as the angles between the three-momenta of $\bar f=\bar \nu,\bar
u,\bar c$, $f'=\mu^-,d,s$ and $\tilde
\chi_1^0$ in the $\tilde \chi_1^-$ rest frame, respectively, and the $\tilde
\chi_1^-$ spin (see Fig.~\ref{fig:5}). Analogous definitions
hold for the $\tilde \chi_1^+$ decay products. Integrating all variables except 
$\theta_{\bar f}$, $\theta_{f'}$ or $\theta_0$, the angular decay
distributions read
\begin{eqnarray}
\frac{1}{\Gamma^{(-)}} \frac{d\Gamma^{(-)}}{d\cos \theta_{\bar f}} & = &
 \frac{1+ h_{\bar f} \cos \theta_{\bar f}}{2} \,, \nonumber \\
\frac{1}{\Gamma^{(-)}} \frac{d\Gamma^{(-)}}{d\cos \theta_{f'}} & = &
 \frac{1+ h_{f'} \cos \theta_{f'}}{2} \,, \nonumber \\
\frac{1}{\Gamma^{(-)}} \frac{d\Gamma^{(-)}}{d\cos \theta_0} & = &
 \frac{1+ h_0^{(-)} \cos \theta_0}{2} \,,
\label{ec:dist}
\end{eqnarray}
with $\Gamma^{(-)} \equiv \Gamma(\tilde \chi_1^- \to \bar f f' \tilde
\chi_1^0)$. The
$h$ factors are called ``spin analysing power'' of the corresponding fermion
$\bar f$, $f'$, $\tilde \chi_1^0$,
and are constants between $-1$ and $1$ which depend on the type of
fermion and the supersymmetric scenario considered. For $\tilde
\chi_1^+$ decays the angular distributions $d\Gamma^{(+)}/d\cos
\theta_{f,\bar{f'},0}$ are given by similar expressions
with constants $h_f$, $h_{\bar f'}$, $h_0^{(+)}$, which
satisfy $h_f=-h_{\bar f}$, $h_{\bar f'}=-h_{f'}$, $h_0^{(+)} = - h_0^{(-)}$
if CP is conserved.

\begin{figure}[htb]
\begin{center}
\epsfig{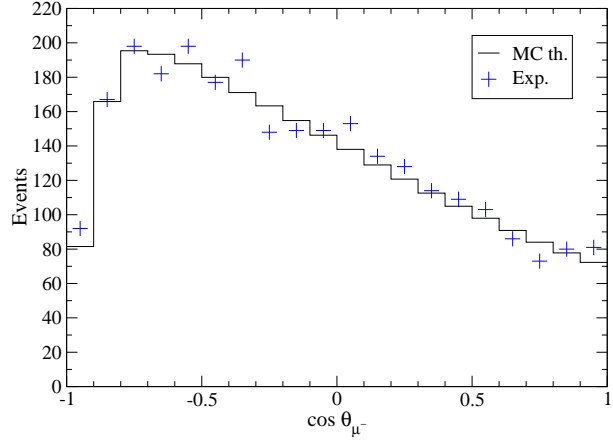}
\caption{Angular distribution of the $\mu^-$ with respect to the $\tilde
\chi_1^-$ spin, in the $\tilde \chi_1^-$ rest frame.}
\label{fig:mum}
\end{center}
\end{figure}

The angular distribution of the muon resulting from $\tilde \chi_1^-$ decay is
shown in Fig.~\ref{fig:mum}. For $\cos \phi_{\mu^-} \simeq -1$
the muon is produced opposite to the $\tilde \chi_1^-$ momentum, then
approximately parallel to
the $e^+$ momentum (up to a Lorentz boost). Then, these events are
suppressed by the requirement of lego-plot separation $\Delta R \geq 0.4$.
With an accurate modelling of the real detector, all the $\cos \theta_{\mu^-}$
range could be eventually included in a fit to experimental data. In our study
we restrict ourselves to the ranges where kinematical cuts do not alter the
distributions. The fit to the data points (discarding the first three bins)
 gives
$h_{\mu^-} = -0.270 \pm 0.016$, in good agreement with the theoretical value
$h_{\mu^-} = -0.252$. 

\begin{figure}[htb]
\begin{center}
\epsfig{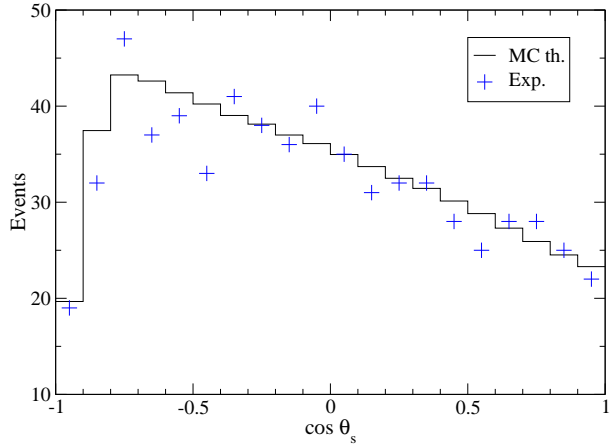}
\caption{Angular distribution of the $s$ quark with respect to the $\tilde
\chi_1^-$ spin, in the $\tilde \chi_1^-$ rest frame.}
\label{fig:s}
\end{center}
\end{figure}

\begin{figure}[htb]
\begin{center}
\epsfig{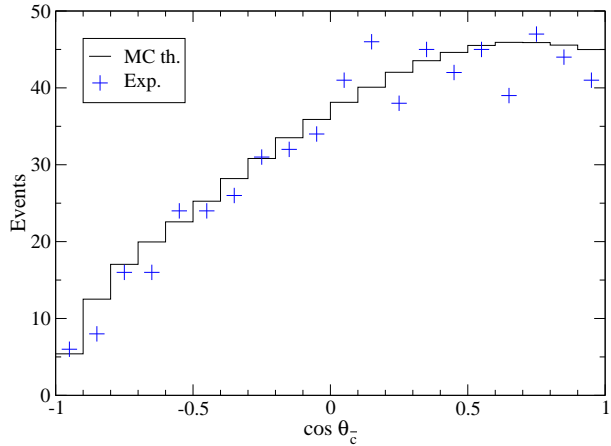}
\caption{Angular distribution of the $\bar c$ antiquark with respect to the
$\tilde \chi_1^-$ spin, in the $\tilde \chi_1^-$ rest frame.}
\label{fig:cbar}
\end{center}
\end{figure}

The study of quark angular distributions requires $c$ tagging to distinguish
between the two quark jets. This reduces the signal by a factor of four, and
thus decreases the statistical accuracy of the measurements. The angular
distribution of the $s$ quark is presented in Fig.~\ref{fig:s}. The suppression
around $\cos \theta_s \simeq -1$ is again caused by the requirement of
lego-plot
separation. The fit to the data points gives $h_s = -0.151 \pm 0.020$, to be
compared with the real value $h_s = -0.149$.
The distribution of the $\bar c$ antiquark is presented in Fig.~\ref{fig:cbar}.
In addition to the suppresion at $\cos \theta_{\bar c} \simeq -1$, we notice a
decrease around $\cos \theta_{\bar c} \sim 1$, which is indirectly caused by
the depression at $\cos \theta_s \simeq -1$. We thus discard the last seven
bins for the fit to the distribution and obtain $h_{\bar c} = 0.387 \pm 0.044$,
being the real value $h_{\bar c} = 0.339$.

As shown by these examples, the ``spin analysing power'' constants governing
the angular
distributions of $\tilde \chi_1^-$ are expected to be measured with an
accuracy which ranges from
6\% for $h_{\mu^-}$ to 13\% for $h_s$. We note that the samples of
``experimental measurements'' generated have been chosen so that the total
number of observed events is around $1\, \sigma$ away from the expected
number, thus the difference between the measured and true values of the $h$
constants is due to statistics. The analysis of $\tilde \chi_1^+$ decays can be
carried out analogously. However, it is interesting to point out that
in CP-conserving scenarios the decays $\tilde \chi_1^- \to \bar f f'
\tilde \chi_1^0$, $\tilde \chi_1^+ \to f \bar f' \tilde \chi_1^0$ can be
summed, substituting $\cos \theta_{\mu^+,c,\bar s} \to -\cos
\theta_{\mu^-,\bar c,s}$, improving
the statistics by a factor $\sim \sqrt 2$. Even in CP-violating scenarios this
is likely to be a good approximation, because $h_{\mu^-}+h_{\mu^+}$,
$h_c+h_{\bar c}$, $h_s+h_{\bar s}$ are negligible at the tree level, as will be
shown in the next subsection, and get nonzero values only through higher order
corrections.

\subsection{CP violation in chargino decays}

The construction of CP-violating observables involves the comparison of $\tilde
\chi_1^-$ and $\tilde \chi_1^+$ decays. Under CP, the vectors in
Fig.~\ref{fig:5} transform to 
\begin{equation}
\vec s_- \to \vec s_+ \,, \quad
\vec p_{\bar f} \to -\vec p_{f} \;,\quad
\vec p_{f'} \to -\vec p_{\bar f'} \;,\quad
\vec p_{\tilde \chi_1^0} \to -\vec p_{\tilde \chi_1^0} \;,
\end{equation}
defined for the $\chi_1^+$ decay.
Then, if we sum $\tilde \chi_1^-$ and $\tilde \chi_1^+$ decays
the products
\begin{eqnarray}
Q_\mu & = & \vec s_\pm \cdot \vec p_{\mu^\pm} \,, \nonumber \\
Q_c & = & \vec s_\pm \cdot \vec p_{c (\bar c)} \,, \nonumber \\
Q_s & = & \vec s_\pm \cdot \vec p_{\bar s (s)}
\end{eqnarray}
are CP-odd. In hadronic decays an additional product $\vec s_\pm \cdot \vec
p_{\tilde \chi_1^0}$ can be measured but it equals $-(Q_c+Q_s)$. Since
these products are even under naive T reversal, the CP-violating asymmetries
\begin{equation}
A_X = \frac{N(Q_X > 0) - N(Q_X < 0)}{N(Q_X > 0) + N(Q_X < 0)} \,,
\end{equation}
with $X=\mu,c,s$ and $N$ standing for the number of events, are also T-even.
Hence, they need the presence of absorptive CP-conserving phases in the
amplitudes in order to be nonvanishing. In the processes under consideration
these phases are generated at the tree level by intermediate particle widths,
but they are extremely small, giving asymmetries $A_X \sim 10^{-3}$. At one
loop level,
these asymmetries can result from the interference between a
dominant tree-level and a subleading one-loop diagram with a CP-conserving
phase, and are expected to be very small as well.

Neglecting for the moment all particle widths (which are kept in all our
numerical computations), the asymmetries $A_X$ vanish at the tree level. Using
the fact that in this approximation the tree-level partial rates
$\Gamma(\tilde \chi_1^- \to \bar f f' \tilde \chi_1^0)$ and 
$\Gamma(\tilde \chi_1^+ \to f \bar f' \tilde \chi_1^0)$ are equal (because
partial rate CP asymmetries are also T-even), these asymmetries are
\begin{equation}
A_X = \frac{h_X+h_{\bar X}}{4} \,.
\end{equation}
Thus, $h_{\mu^-}+h_{\mu^+}$, $h_c+h_{\bar c}$, $h_s+h_{\bar s}$ vanish at the
tree level if particle widths are neglected, and take values $O(10^{-3})$ (much
smaller than the precision of the measurement of the individual constants) when
particle widths are included. 

A CP-odd, T-odd asymmetry can be built from the product
\begin{equation}
Q_{12} = \vec s_\pm \cdot \left( \vec p_{\bar q_1} \times \vec p_{q_2}
\right) \,, 
\end{equation}
where $\vec p_{\bar q_1}$ and $\vec p_{q_2}$ are the momenta of the
antiquark and quark resulting from the hadronic $\tilde \chi_1^\pm$ decay,
respectively. (A triple
product $\vec s_\pm \cdot ( \vec p_{X_1} \times \vec p_{X_2})$, with $X_1=c,
\bar c$, $X_2 = \bar s,s$ distinguished by quark flavour instead of baryon
number, is CP-even.) The quark and antiquark are distinguished using $c$ tagging
and looking at the charge of the muon, as explained in section \ref{sec:3}.
The CP asymmetry
\begin{equation}
A_{12} = \frac{N(Q_{12} > 0) - N(Q_{12} < 0)}{N(Q_{12} > 0) + N(Q_{12} < 0)}
\end{equation}
can be sizeable already at the tree level, and without the need of interference
between different Feynman diagrams. As discussed in section \ref{sec:2},
in the SUSY scenario selected the $\tilde
\chi_1^-$ decays are strongly dominated by diagram (a) in Fig.~\ref{fig:3}.
A large triple-product CP asymmetry is possible because the polarised decay
width for
$\tilde \chi_1^- \to W^- \tilde \chi_1^0 \to \bar f f' \tilde \chi_1^0$
contains a term proportional to
$\epsilon_{\mu \nu \rho \sigma} s_-^\mu \, P_{\bar \chi_1^-}^\nu \,
P_{\bar f}^\rho \, P_{f'}^\sigma \; m_{\tilde \chi_1^0} \; \mathrm{Im} \,
O_R^{11} O_L^{11*}$,
where $O_{L,R}^{11}$ denote the left- and right-handed parts of the (complex)
$W \tilde \chi_1^0 \tilde \chi_1^-$ coupling, respectively, and the momenta
follow obvious notation. In the $\tilde \chi_1^-$ rest frame
we have $P_{\bar \chi_1^-} = ( \, m_{\tilde \chi_1^-},\vec 0 \, )$,
$\vec P_{\bar f} = \vec p_{\bar f}$, $\vec P_{f'} = \vec p_{f'}$ and this term
reduces to a triple product.

\begin{figure}[htb]
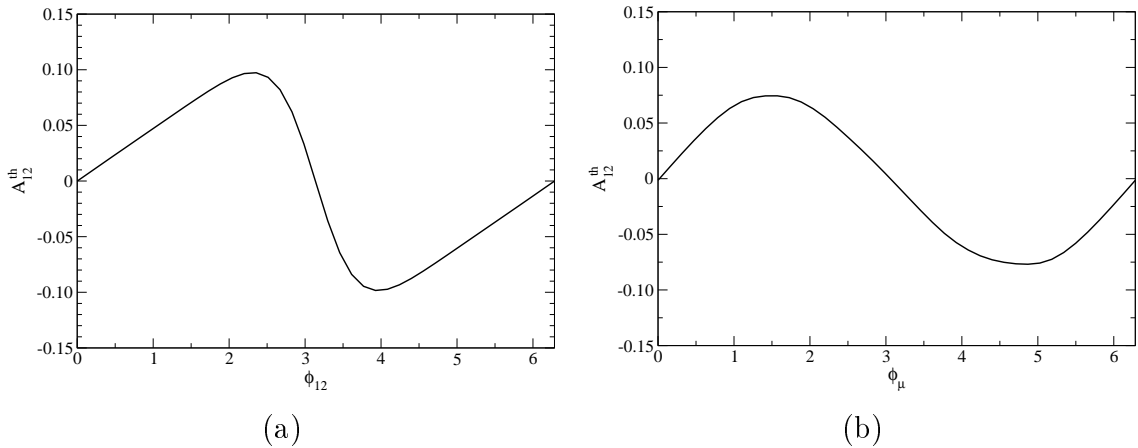

\begin{center}
\begin{tabular}{cc}
\epsfig{file=Figs/asim-th.eps,width=7.3cm,clip=} &
\epsfig{file=Figs/asim-mu-th.eps,width=7.3cm,clip=} \\
(a) & (b)
\end{tabular}
\caption{Theoretical value of $A_{12}$ as a function of $\phi_1$, for
$\phi_\mu =0$ (a) and as a function of $\phi_\mu$, for $\phi_1=0$ (b).}
\label{fig:A12th}
\end{center}
\end{figure}

The theoretical value of $A_{12}$ as a function of  $\phi_1$, taking
$\phi_\mu =0$, is shown in Fig.~\ref{fig:A12th} (a), where we observe that
the asymmetry can reach
values $O(0.1)$ for large phases $\phi_1 \simeq 2$. For large $\phi_\mu$ values
the asymmetry could be large too, as can be seen
in  Fig.~\ref{fig:A12th} (b). However,
for the range $|\phi_\mu| \lesssim 0.12$ (modulo $\pi$) allowed by EDM constraints
the variation of $A_{12}$ is only between $\pm 0.01$. 

\begin{figure}[ht]
\begin{center}
\epsfig{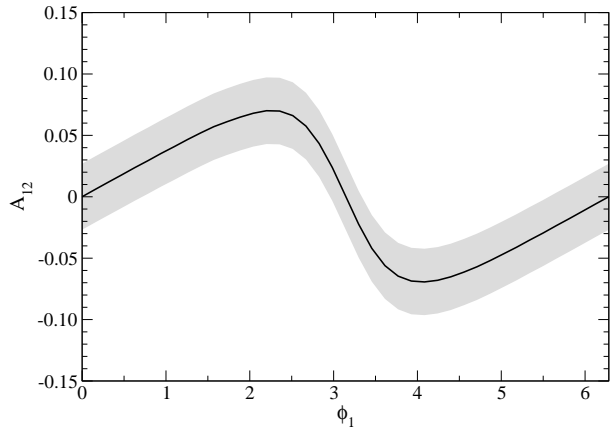}
\caption{Asymmetry $A_{12}$ as a function of the phase $\phi_1$, for
$\phi_\mu$ values fulfilling EDM constraints, as explained in the text. The
shaded region represents the statistical error for one year of running.}
\label{fig:A12}
\end{center}
\end{figure}

We present in Fig.~\ref{fig:A12} the value of the asymmetry after signal
reconstruction, including the corrections from ISR, beamstrahlung, etc.
discussed in section~\ref{sec:3}. The shaded region
represents the statistical error for one year of running with an integrated
luminosity of 534 fb$^{-1}$. The variation of the cross section with $\phi_1$ is
not relevant. In this plot we have chosen pairs of phases $(\phi_1,\phi_\mu)$
allowed by EDM constraints: for each $\phi_1$, we take a
$\phi_\mu$ value, with $|\phi_\mu| \lesssim 0.12$, for which the neutralino and
chargino contributions to the electron EDM cancel. Then, we calculate the
asymmetry for this $(\phi_1,\phi_\mu)$ pair. (Several representative examples of
these allowed pairs of phases are collected in Table~\ref{tab:pairs}.)
The maximum differences in the asymmetry between taking $\phi_\mu=0$ and taking
the $\phi_\mu$ values required by EDM constraints are of 10\%, found for
$\phi_1 \sim 0,\pi$.

\section{Conclusions}
\label{sec:5}

In this paper we have shown that, in SUSY scenarios where the decays
$\tilde \nu_e \to e^- \tilde \chi_1^+$, $\tilde \nu_e^* \to e^+ \tilde \chi_1^-$
are kinematically allowed, sneutrino pair production provides a copious source
of 100\% polarised charginos, in which the kinematics of the process allows to
reconstruct their four-momenta. The large cross section for the full process
$e^+ e^-  \to \tilde \nu_e^* \tilde \nu_e \to e^+ \tilde \chi_1^- \,
e^- \tilde \chi_1^+ \to e^+ \bar \nu_\mu \mu^- \tilde \chi_1^0 \,
e^- q \bar q' \tilde \chi_1^0$ (plus its charge conjugate) and the low
backgrounds allow precise measurements of spin-related quantities like angular
distributions and triple product CP-violating asymmetries.

The $e^-$ distribution in the $\tilde \nu_e$ rest frame shows that the $\tilde
\nu_e$ decay is isotropic, what is a strong indication that sneutrinos are
scalars. Since the spin of electrons is $1/2$, total angular momentum
conservation implies that the charginos have half-integer spin, being 
$1/2$ the most obvious possibility. Angular
distributions in $\tilde \chi_1^-$ rest frame allow to measure the values of the
``spin analysing power'' constants $h_{\mu^-}$, $h_{\bar c}$, $h_s$ controlling
the
angular distributions of its decay products, with a precision between 6\% and
13\% for only one year of running. Since the sums $h_{\mu^-}+h_{\mu^+}$,
$h_c+h_{\bar c}$, $h_s+h_{\bar s}$ are zero if CP is conserved, and are expected
to be small in CP-violating scenarios (even including radiative corrections),
the data from $\tilde \chi_1^+$ decays could eventually be included in the
analyses, improving the statistical precision.

\begin{figure}[htb]
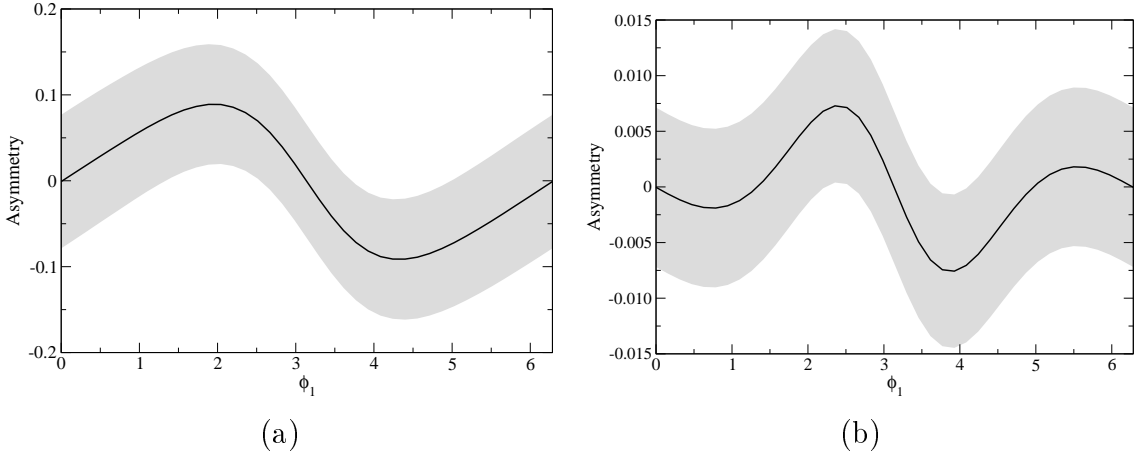

\begin{center}
\begin{tabular}{cc}
\epsfig{file=Figs/asim-casc.eps,width=7.3cm,clip=} &
\epsfig{file=Figs/asim-xx.eps,width=7.3cm,clip=} \\
(a) & (b)
\end{tabular}
\caption{Triple product CP-violating asymmetries
defined in (a)
selectron cascade decays \cite{casc}; (b) $\tilde \chi_1^0 \tilde \chi_2^0$
production \cite{x1x2}, in the same SUSY scenario studied here. The shaded
regions represent the statistical error for one year of running.}
\label{fig:A1}
\end{center}
\end{figure}

Despite the fact that CP-violating phases could be detected in CP-conserving
quantities \cite{CPcon}, the direct observation of supersymmetric CP violation
is extremely important. We have shown that
in $\tilde \chi_1^\pm$ hadronic decays the CP-violating asymmetry in the
product $\vec s_\pm \cdot (\vec p_{\bar q_1} \times \vec p_{q_2})$ is very
sensitive to the phase of $M_1$, and can have
values up to $A_{12} = 0.07$ for $\phi_1 \simeq 2$. Such asymmetry could be
observed with $2.6 \, \sigma$ statistical significance in one year of running.
It is worth comparing
this sensitivity with the ones obtained in other processes
within the same SUSY scenario. In selectron cascade decays
$\tilde e_L \to e \tilde \chi_2^0 \to e \tilde \chi_1^0 \mu^+ \mu^-$, a
CP asymmetry in the triple product
$\vec s \cdot (\vec p_{\mu^-} \times \vec p_{\mu^+})$ can be built,
with $\vec s$ the $\tilde \chi_2^0$ spin \cite{casc}. In $\tilde \chi_1^0
\tilde \chi_2^0$ production $e^+ e^- \to \tilde \chi_1^0 \tilde \chi_2^0 \to
\tilde \chi_1^0 \tilde \chi_1^0 \ell^+  \ell^-$, $\ell=e,\mu$, an analogous
 asymmetry in the
product  $\vec p_{e^+} \cdot ( \vec p_{\ell^-} \times \vec p_{\ell^+} )$ can be
defined \cite{x1x2,x1x2other}. The values of these asymmetries as a function of
$\phi_1$  (including ISR, beamstrahlung, beam spread and detector effects, as
well as backgrounds)  are shown in Fig.~\ref{fig:A1} (adapted from
Refs.~\cite{casc,x1x2}, where two years of running are considered instead of
one). Comparing Figs.~\ref{fig:A12},~\ref{fig:A1} 
it is apparent that the observation of CP violating effects is much easier in
sneutrino cascade decays. For large phases $\phi_1 \simeq 2$ the statistical
significance of the CP asymmetry $A_{12}$ is a factor of two larger than for
the asymmetries in selectron cascade decays and neutralino pair production.
Therefore, provided a good $c$ tagging efficiency is achieved, sneutrino
cascade decays provide a much more sensitive tool to test CP violation in the
neutralino sector than neutralino production and decay proceses.

\vspace{1cm}
\noindent
{\Large \bf Acknowledgements}

\vspace{0.4cm} \noindent
This work has been supported
by the European Community's Human Potential Programme under contract
HTRN--CT--2000--00149 Physics at Colliders and by FCT
through projects POCTI/FNU/43793/2002, CFIF--Plurianual (2/91) and
grant SFRH/ BPD/12603/2003.

\end{document}